\documentclass[referee]{raa}            

\usepackage{graphicx,times}             
\usepackage{natbib}
\usepackage{amssymb,amsmath}
\bibpunct{(}{)}{;}{a}{}{,}

\usepackage[a4paper=true,dvipdfm=true,pagebackref=true]{hyperref}
\hypersetup{colorlinks = true, linkcolor = green, anchorcolor = red, citecolor = blue, filecolor = red, pagecolor = red, urlcolor = red}

\begin{document}

   \title{Constraining Evolution of Magnetic Field Strength in Dissipation Region of Two BL Lac Objects
}

   \volnopage{Vol.0 (20xx) No.0, 000--000}      
   \setcounter{page}{1}          

   \author{Xu-Liang Fan
      \inst{1}
   \and Da-Hai Yan
      \inst{2,3}
   \and Qingwen Wu
      \inst{4}
   \and Xu Chen
      \inst{5}
   }

   \institute{School of Mathematics, Physics and Statistics, Shanghai University of Engineering Science, Shanghai 201620, China; {\it fanxl@sues.edu.cn}\\
        \and
             Yunnan Observatories, Chinese Academy of Sciences, Kunming 650011, China
        \and
             Key Laboratory for the Structure and Evolution of Celestial Objects, Chinese Academy of Sciences, Kunming 650011, China\\
        \and
             School of Physics, Huazhong University of Science and Technology, Wuhan 430074, China
        \and
             Shandong Provincial Key Laboratory of Optical Astronomy and Solar-Terrestrial Environment, Institute of Space Sciences, Shandong University, Weihai, 264209, China\\
\vs\no
   {\small Received~~20xx month day; accepted~~20xx~~month day}}

\abstract{With the assumption that the optical variability timescale is dominated by the cooling time of the synchrotron process for BL Lac objects, we estimate time dependent magnetic field strength of the emission region for two BL Lac objects. The average magnetic field strengths are consistent with those estimated from core shift measurement and spectral energy distribution modelling. Variation of magnetic field strength in dissipation region is discovered. Variability of flux and magnetic field strength show no clear correlation, which indicates the variation of magnetic field is not the dominant reason of variability origin. The evolution of magnetic field strength can provide another approach to constrain the energy dissipation mechanism in jet.
}

   \authorrunning{X.-L Fan et al.}            
   \titlerunning{Magnetic Field of BL Lacs}  

   \maketitle

%
%
\section{Introduction}           
\label{sect:intro}

Magnetic field and magnetization (characterized by magnetization parameter $\sigma$) inside jets are important to understand acceleration and energy dissipation mechanisms of relativistic jets~\citep{2010MNRAS.402..353L, 2013ApJ...764L..24S, 2019ARA&A..57..467B}. The flux variability of jet is also suggested to be related to magnetic reconnection process~\citep{2013MNRAS.431..355G, 2018ApJ...856...80F, 2020NatCo..11.4176S}. The methods to constrain magnetic field structure in jet are usually based on the polarization measurements~\citep{2012AJ....144..105H, 2018ApJ...862..151H}. Inside the emission zone of blazar, which is believed to be dominated by pc-scale jet, magnetic field strength can be estimated from modelling of multi-wavelength spectral energy distribution (SED)~\citep{2014ApJ...788..104Z, 2018ApJS..235...39C} and core shift measurements (with equipartition assumption,~\citealt{2012A&A...545A.113P,2014Natur.510..126Z} or without equipartition assumption,~\citealt{2015MNRAS.451..927Z}).

According to the results of core shift measurement and several theoretical arguments, magnetic field strength decreases along jet, with $B \propto r^{-1}$, where $r$ is the distance between jet and the central engine~\citep{2009MNRAS.400...26O, 1979ApJ...232...34B, 2021ApJ...906..105C}. Based on this relation, magnetic field strength inside the dissipation region was suggested to constrain the location of emission region~\citep{2018ApJ...852...45W, 2018ApJ...859..168Y}.

Similar with the electromagnetic radiation, the strength of magnetic field strength is also suggested as variable over time (e.g.,~\citealt{2011MNRAS.410..368B, 2019Galax...7...35T, 2021MNRAS.505.6103P}). The variation of magnetic field strength could also be one possible origin of flux variability of blazars~\citep{2011ApJ...736..128P}. The results of core shift measurements showed that magnetic field strength was variable during the flux flares, which was possibly related to the new jet component~\citep{2019MNRAS.485.1822P}. The direction of magnetic field in jet is also found to be variable~\citep{2018ApJ...862..151H}.

Similar to the method based on SED modelling, discovering variation of magnetic field strength from core shift measurement is based on the flux measurement~\citep{2019MNRAS.485.1822P}. However, the magnetic field strength estimated by these two methods can differ by a factor of 3~\citep{2014ApJ...796L...5N}. There was also suggestions that biases from core shift measurements can overestimate magnetic field strength in jet~\citep{2020MNRAS.499.4515P}. Thus there needs an independent method to estimate the magnetic field strength and its evolution inside the dissipation region. If the optical variability are mainly dominated by the cooling from synchrotron radiation, the lower limit of the magnetic field strength can be constrained with the variability timescale~\citep{2003ApJ...596..847B}. In this paper, we constrain the optical variability timescale of two BL Lac objects (BL Lacs) with high sampling intra-day observations. Then we estimate the magnetic field strength of their emission regions and explore its evolution at timescales of years. In Section 2, we give the method to estimate magnetic field strength based on the optical photometric data, and the results of estimated variability timescale and magnetic field strength. Comparison of magnetic field strength with results derived from other methods and implications for evolution of magnetic field strength are discussed in Section 3. Section 4 summarizes the main conclusions.


\section{Method and Results}
\label{sect:method}
The SEDs of blazars are dominated by the non-thermal radiation of jet, which show two bumps on the $\nu$ --- $\nu F_{\nu}$ diagram~\citep{2010ApJ...716...30A}. The low energy bump is believed to be produced by synchrotron emission. Synchrotron self-Compton (SSC) is suggested as the dominant mechanism for the high energy bump of BL Lacs --- a subclasses of blazars with weak emission lines, especially high energy peaked BL Lacs (HBLs)~\citep{2010ApJ...716...30A}.

The intra-day or micro variability at optical band is a characteristic property of BL Lacs~\citep{1995ARA&A..33..163W}. The timescale of the intra-day variability can be as short as several minutes (e.g., \citealt{1998A&AS..132...83B, 2009ApJS..181..466F, 2014MNRAS.443.2940H}). For BL Lacs, the optical emission is generally dominated by the synchrotron radiation, combined with possible host starlight for nearby sources. Thus, by ignoring cooling from inverse Compton scattering, the typical variability timescale can be seen as the upper limit of the cooling time of synchrotron radiation.

The cooling time of the synchrotron radiation can be estimated by~\citep{1998ApJ...509..608T, 2003ApJ...596..847B}
\begin{equation}
 t_{cool} = \frac{3}{4} \frac{m_ec^2}{\sigma_T c}(\gamma u_B)^{-1}
          = \frac{6\pi m_ec}{\sigma_T\gamma B^2}~~{\rm s},
\label{tcool}
\end{equation}
where $u_B = B^2/8\pi$ is the energy density of the magnetic field, $m_e$ is the mass of electron, $\sigma_T$ is the cross section of Thomson scattering, $\gamma$ is the electron energy. Meanwhile, the observational frequency is related to the electron energy $\gamma$ with
\begin{equation}
 \nu = \frac{4}{3} \nu_L \gamma^2 \frac{\delta}{1+z}
     = 3.7\times10^6 \gamma^2 B \frac{\delta}{1+z}~~{\rm Hz},
\label{nu}
\end{equation}
where $\nu_L = 2.8\times10^6 B$ is the Larmor frequency, $\delta$ is the Doppler factor, and $z$ is redshift.

Considering the observational variability timescale as the upper limit of $t_{cool}$, i.e., $t_{var}\delta /(1+z) \geq t_{cool}$, one can get the lower limit of the magnetic field strength combined equation~\ref{tcool} and~\ref{nu},
\begin{equation}
 B \geq 1.31\times 10^8 t_{var}^{-2/3} \nu^{-1/3} \delta^{-1/3} (1+z)^{1/3}~~{\rm G},
\label{estimatemag}
\end{equation}
where $t_{var}$ in unit of second and $\nu$ in unit of Hz. B is in unit of Gauss.

\begin{figure}
   \centering
   \includegraphics[width=\textwidth, angle=0]{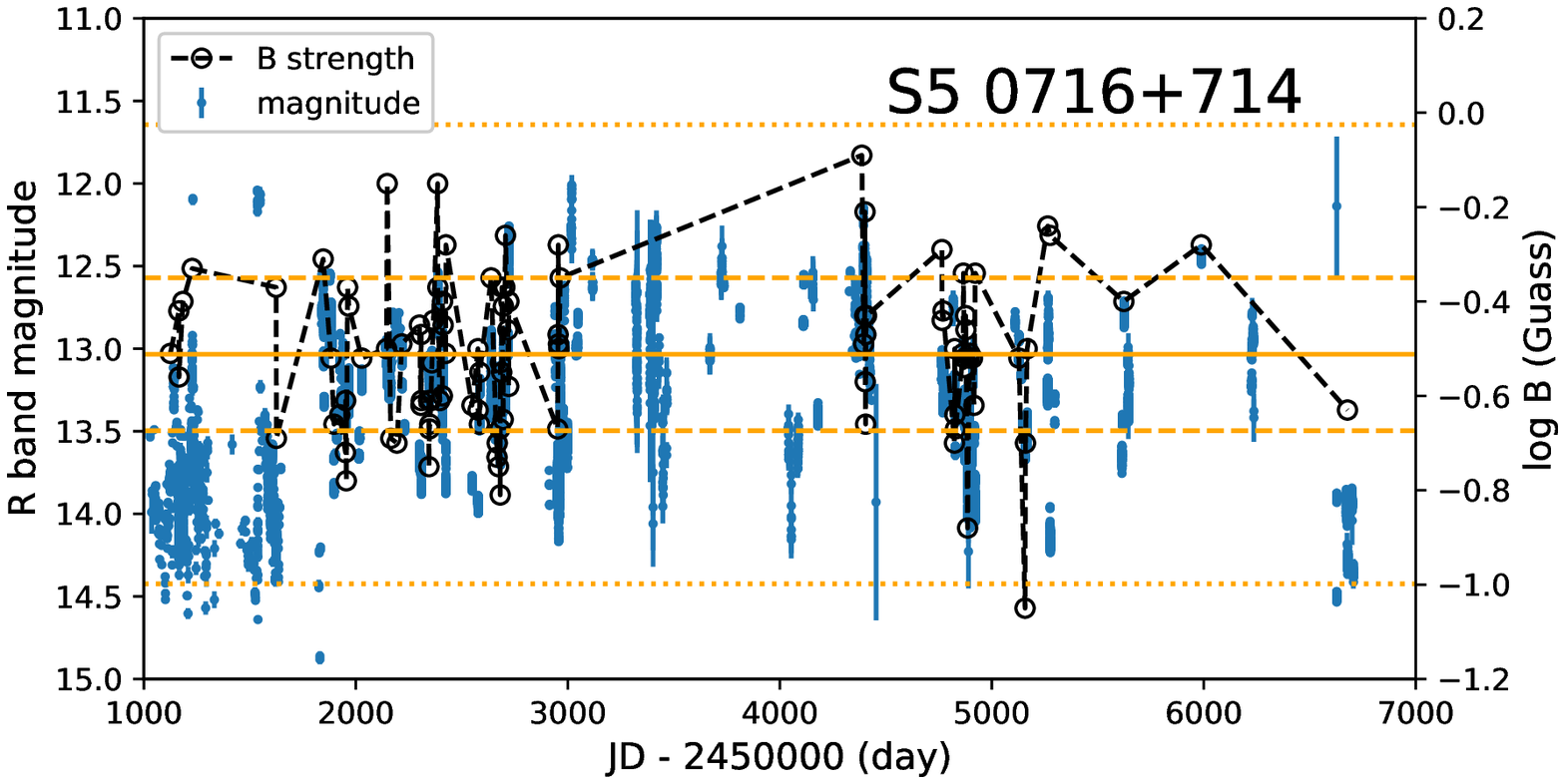}
   \includegraphics[width=\textwidth, angle=0]{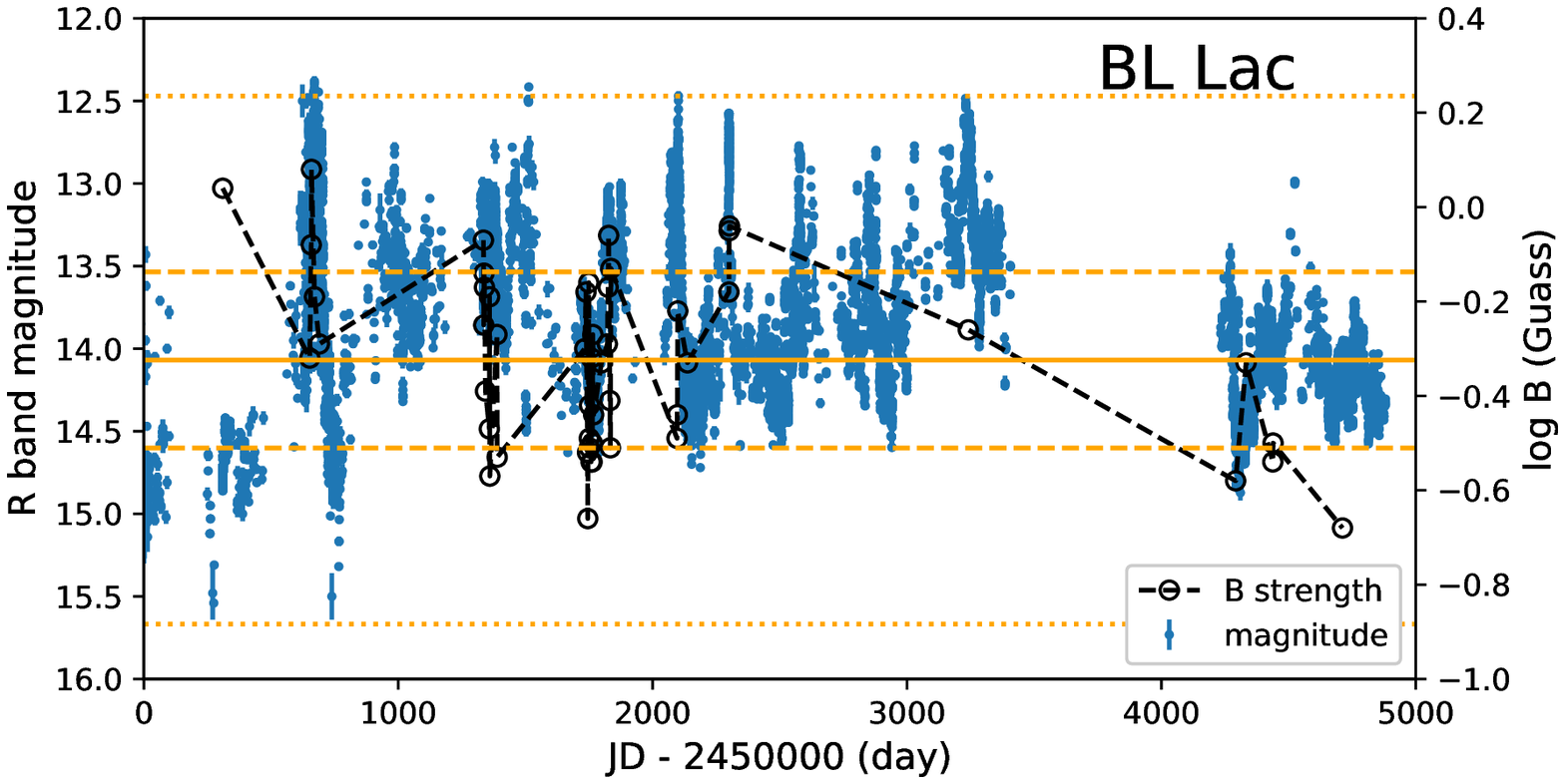}
   \caption{The variations of lower limit of magnetic field strength and R band magnitude for S5 0716+714 (Upper panel) and BL Lacertae (Lower panel). The orange solid lines show the mean values of magnetic field strength. The dashed and dotted lines represent 1$\sigma$ and 3$\sigma$ for the distribution of $\log B$, respectively.\label{mag}}
   \end{figure}

Based on Equation~\ref{estimatemag}, if the observational variability timescale is obtained for a blazar, one can estimate the lower limit of the magnetic field strength with a special Doppler factor (taken as 10 throughout this paper). Our purpose in this work is to estimate the magnetic field strength with the intra-day variability timescale and explore its possible evolution. Therefore, we search the literature between 1990 and 2016 for the historical lightcurve at R band ($\nu = 4.6769\times10^{14}$ Hz). There are two sources (S5 0716+714 and BL Lacertae) with relatively more data and higher data sampling to derive the variability timescale on days or even hours (the references are list in Table~\ref{ref}). The lightcurves of both sources are shown in Figure~\ref{mag}. For both sources, the lightcurve is divided into intra-day timescales according to the observed time (Julian day). If the time interval of two adjacent data points is longer than two hours, we just divide them into two lightcurves. Otherwise they will be considered to belong to a single lightcurve. This criterion is for the relatively continuous lightcurve without large time gaps, and it will not divide the lightcurves across two Julian days. In order to estimate the variability timescale better, we intend to choose the intra-day lightcurves with relatively better sampling, longer lasting observed time, as well as obvious variability. Thus, the lightcuvres with observed time spanning longer than 0.5 hour, number of data points larger than 20, and magnitude varying more than $5 \sigma$ (where $\sigma$ is the observational errors) are retained. Then the lightcurves are inspected by eyes to exclude the ones with random variations caused by the weather conditions or instrumental reasons. After these selection criteria, the number of intra-day lightcurves is significantly reduced. For the remaining lightcurves, we estimate the variability timescale of each lightcurve with~\citep{1995ARA&A..33..163W}
\begin{equation}
 \tau = \frac{<F>}{\mid \Delta F/\Delta t \mid }
\label{tau}
\end{equation}
where $<F>$ is the average flux during the observational time range of individual lightcurve, $\Delta F$ is the flux difference between the maximum and minimum flux, $\Delta t$ is the time between the maximum and minimum flux.

The variability timescales range from $3.47\times10^{4} s$ to $9.12\times10^{5} s$ for S5 0716+714, and from $1.74\times10^{4} s$ to $2.34\times10^{5} s$ for BL Lacertae. Based on the variability timescales, we estimate the lower limit of the magnetic field strength for the selected time range with Equation~\ref{estimatemag} (z is taken as 0.3 and 0.0686 for S5 0716+714 and BL Lacertae, respectively). The variations of the magnetic field over time for both sources are plotted in Figure~\ref{mag}

\begin{table}
\begin{center}
  \caption[]{The references for historical data}
  \label{ref}
  \begin{tabular}{cc}
  \hline
  Objects & Reference \\
  \hline
  S5 0716+714 & Q02, R03, G06, M06, G08, Z08, P09, C11, B13, H14, D15, WEBT (V00, V08, O06) \\
  BL Lac & B98, B99, X99, F00, F01, C01, H04, Z04, G06, WEBT (V09, R09, R10) \\
  \hline
  \end{tabular}
  \tablecomments{0.88\textwidth}{
  B98:~\citet{1998A&AS..132...83B}, B99:~\citet{1999A&AS..136..455B}, X99:~\citet{1999ApJ...522..846X}, F00:~\citet{2000ApJ...537..101F}, V00:~\citet{2000A&A...363..108V}, C01:~\citet{2001AJ....121...90C}, F01:~\citet{2001A&A...369..758F}, Q02:~\citet{2002AJ....123..678Q}, R03:~\citet{2003A&A...402..151R}, H04:~\citet{2004AstL...30..209H}, Z04:~\citet{2004AJ....128.1929Z}, G06:~\citet{2006A&A...450...39G}, M06:~\citet{2006A&A...451..435M}, O06:~\citet{2006A&A...451..797O}, G08:~\citet{2008AJ....135.1384G}, V08:~\citet{2008A&A...481L..79V}, Z08:~\citet{2008AJ....136.1846Z}, P09:~\citet{2009ApJS..185..511P}, R09:~\citet{2009A&A...507..769R}, V09:~\citet{2009A&A...501..455V}, R10:~\citet{2010A&A...524A..43R}, C11:~\citet{2011ApJ...731..118C}, B13:~\citet{2013A&A...558A..92B}, H14:~\citet{2014MNRAS.443.2940H}, D15:~\citet{2015ApJS..218...18D}
  }
\end{center}
\end{table}

The maximum magnetic field strengthes for S5 0716+714 and BL Lacertae are $\log B = -0.09$ and 0.08, respectively, while the minimum ones are -1.05 and -0.68, respectively. The mean values and standard deviation of $\log B$ are -0.51 and 0.16 for S5 0716+714, which are -0.32 and 0.19 for BL Lacertae. Both S5 0716+714 and BL Lacertae show obvious variations on magnetic field strength ($\Delta B > 3\sigma_{B}$).

%

\section{Discussion}
\label{sect:discussion}
\subsection{The strength of magnetic field inside emission region}
There are several methods to estimate magnetic field strength in jet. The most widely used method for emission region of blazar is based on the SED modelling.~\citet{2020ApJ...898...48A} modelled the SED of these two sources with the quasi-simultaneous multi-frequency data. They derive $\log B = -0.33$ and -0.08 for S5 0716+714 and BL Lacertae, respectively.~\citet{2018ApJS..235...39C} also estimated the parameter of radiation for a sample of Fermi blazars with approximate analytical expressions. The magnetic field strengths $\log B = $ -1.05 and 0.64 are derived for S5 0716+714 and BL Lacertae, respectively.

Core shift measurement provides another method to estimate magnetic field strength and electron number density in jet. The standard method assumes equipartition between the energy of magnetic field and particle, and estimates the magnetic field strength at 1 pc from jet vertex, which further infers the B strength of radio core with $B \propto r^{-1}$~\citep{2009MNRAS.400...26O}. This method gives that $\log B_{1pc} = -0.31$  and -1.05 for S5 0716+714 and BL Lacertae, respectively~\citep{2012A&A...545A.113P}. For the 15 GHz core, $\log B = -1.15$ and -0.96 for S5 0716+714 and BL Lacertae, respectively~\citep{2012A&A...545A.113P}.

\begin{figure*}
\centering
\includegraphics[angle=0,scale=.8]{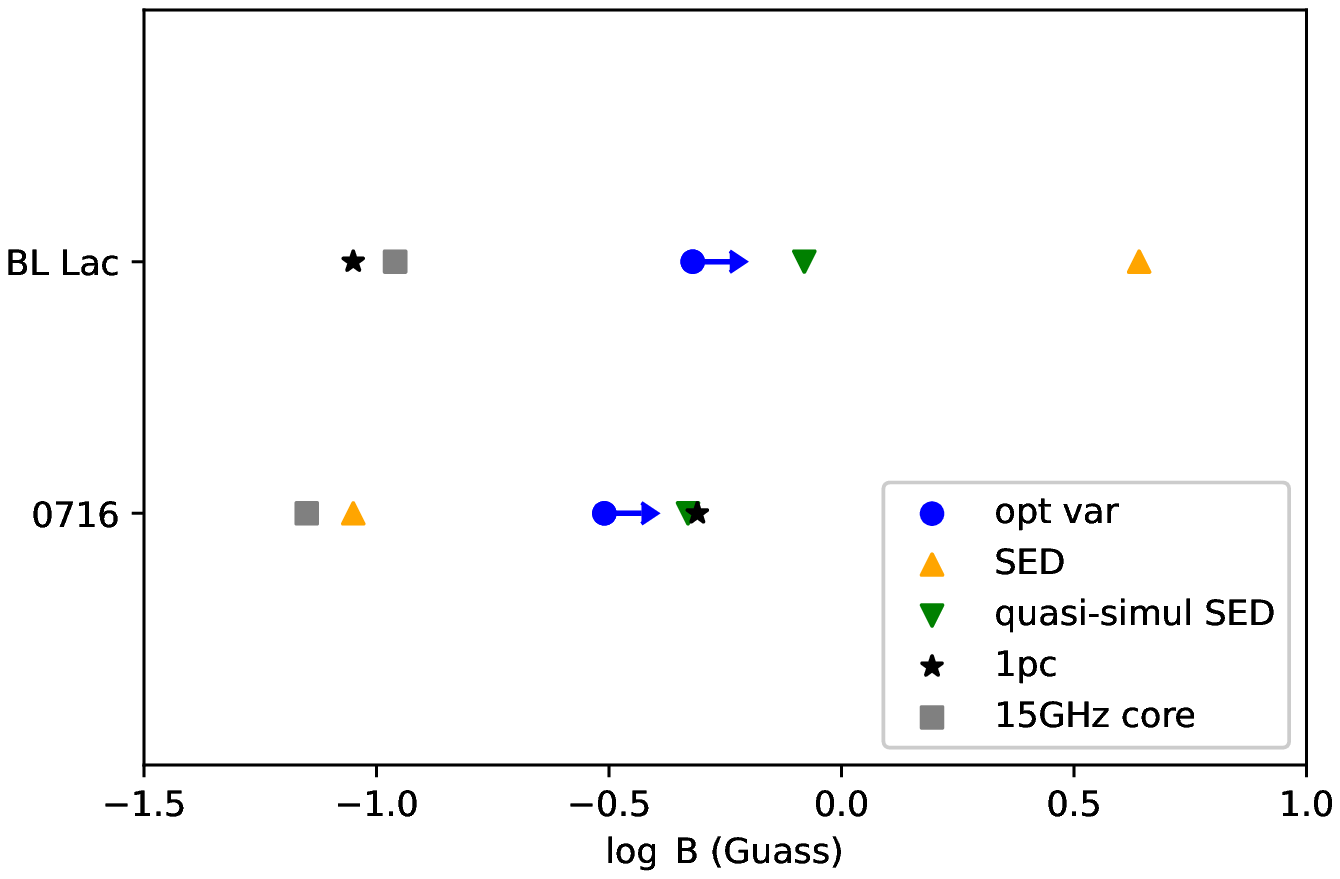}
\caption{Comparison of the estimated magnetic field strength with different methods. The blue circles with arrows represent the mean values of the lower limit of magnetic field strength estimated in this work. See the text for details. \label{mag_com}}
\end{figure*}

The deviation of magnetic fields strengths estimated from different methods above can be as large as 1.7 dex.~\citet{2014ApJ...796L...5N} attempted to reconcile the magnetic dominated jet and the high Compton dominance for flat spectrum radio quasars (FSRQs), as well as the higher magnetic field strengths estimated from core shift measurements. They suggested inhomogeneous magnetic field structure in jet as a potential explanation. The emission region has lower local magnetic field strength compared to the overall jet, due to magnetic reconnection layers or jet spines. In fact, the magnetic field strengths estimated from core shift measurement are not always higher than that from SED modelling, e.g., 0.9 G given by~\citet{2012A&A...545A.113P} versus 1.56 G given by~\citet{2018ApJS..235...39C} for FSRQs. In Figure~\ref{mag_com}, all magnetic field strengths of S5 0716+714 and BL Lacertae listed above are plotted, as well as the mean values of those estimated from optical variability in this work. Generally, magnetic field strengths estimated by this work are at the middle of the other two methods. Considering the large uncertainty in various methods, the magnetic field strength estimated with optical variability is consistent with the results of other estimations.

In section 2, we estimate the lower limit of magnetic field strength with a certain value of Doppler factor (taken as 10). The value of $\log B$ would increase by 0.10 and decrease by 0.23 when the Doppler factor taken as 5 and 50, respectively. Even the most extreme values are considered, the estimated magnetic field strengths still locate at the middle region in Figure~\ref{mag_com}. Another possibility is that Doppler factor is also variable along time. This scenario is suggested by~\citet{2017Natur.552..374R} to explain the long-term variability of blazars (also see~\citealt{2021MNRAS.501.1100R}). The varied Doppler factors will result in varied observational variability timescale. Then the estimated magnetic field strength will also \textbf{vary} according to the variation of Doppler factor. Our results show no obvious long-term trend on the variation of magnetic field strength (Figure~\ref{mag}). However, the variation of estimated magnetic field strength is caused by variation of Doppler factor can not be excluded. More investigations and independent constraints for the variation of Doppler factor are needed in the future.

Magnetic field strength measurement is important to clarify the radiation mechanism of blazars, as strong magnetic filed (about three orders of magnitude higher than that of leptonic model) is required for hadronic model, especially for proton synchrotron emission (e.g.,~\citealt{2019NewAR..8701541H, 2020Galax...8...72C}). The magnetic field $\lesssim 1$ G estimated by all three methods above disfavors proton synchrotron emission for high energy emission, while leptonic or lepto-hadronic model can be compatible with it~\citep{2020Galax...8...72C}.

\citet{2018ApJ...859..168Y} proposed a method to locate the emission region with the estimation of magnetic field strength. If the magnetic field strength estimated from optical variability and core shift are both correct, combined with the relation $B \propto r^{-1}$, one can constrain the location of emission zone. This method requires a precondition that  magnetic field strength and location of emission region should be stable along time. Once B strength in the emission region can be variable, simultaneous measurements are needed for the application of the relation of decreasing B strength with distance.

\subsection{The evolution of magnetic field strength}
As discussed above, under the assumption that the B strength is stable at special location in jet, the strength of B can be used to constrain the location of dissipation region. If this assumption is true at the timescale of years, the observational evolution of B strength means the variation of emission region in jets. As the expectation of adiabatic expansion of jet, the blob would move outward from jet base, which results in decrease of B strength along time. No such continuous behaviours are found in Figure~\ref{mag}.

On the opposite, variable magnetic field along time can also cause the flux variability. Figure~\ref{mag} shows the variation of magnetic field strength of two BL Lacs, as well as their flux variability. No clear trend is found between variability of flux and B strength. This indicates that variation of magnetic field strength in emission region is not the dominant reason of flux variability. The variability origin is not only related to the variation of magnetic field strength, but also other factors. The variation of magnetic field strength can be caused by new jet components as suggested by~\citet{2019MNRAS.485.1822P}, or by turbulence components~\citep{2021Galax...9...27M}.

Polarization observation is an important tool to constrain structure of magnetic field in jet~\citep{2019NewAR..8701541H}. Intra-night variability of polarization degree and position angle was also found for S5 0716+714 and BL Lacertae~\citep{2016ApJ...831...92B, 2020ApJ...900..137W, 2021Galax...9...27M}. The results of polarization behaviors indicated superposition of different turbulent regions or magnetic reconnection.~\citet{2021Galax...9...27M} demonstrated results of their multi-band flux and polarization monitoring for several blazars (including S5 0716+714 and BL Lacertae in this work). They compared the observed results with the predictions of Turbulent Extreme Multi-Zone model, and concluded that disordered magnetic field is important to produce the observed polarization behaviors. To distinguish different energy dissipation processes, combined the polarization variability with the variation of magnetic field strength before and after the flares could be useful. The magnetic field strength is expected to increase before and after shock regions, while it would decrease at the downstream of magnetic reconnection due to conversion from magnetic energy to kinetic energy of radiative particles (e.g.,~\citealt{2010RvMP...82..603Y, 2015MNRAS.450..183S}). Our method provide an approach to estimate magnetic field strength with continuous, high cadence optical observations.

\section{Conclusions}
\label{sect:conclusion}
In this paper, we estimate the optical variability timescale of two BL Lacs with high sampling intra-day lightcurve. Under the assumption that the cooling timescale is dominated by the synchrotron radiation for BL Lacs, we estimate the lower limit of magnetic field strength inside the emission zone. The similar results compared with other methods prove the validity of this method. More importantly, this method can constrain the evolution of magnetic field strength along time. Our results give an independent evidence that the magnetic field strength is variable in the dissipation region of jet. Works with better sampling data and combining polarization observations should give more constraints on the variability origin and energy dissipation mechanisms.

\begin{acknowledgements}
We thank the anonymous referee for the useful comments. This work is supported by National Natural Science Foundation of China (NSFC; grant number 11803081, 11947099, U1931203, and 12003014). The work of D. H. Yan is also supported by the CAS Youth Innovation Promotion Association and Basic research Program of Yunnan Province (202001AW070013). Fractional of this work is based on data taken and assembled by the WEBT collaboration and stored in the WEBT archive at the Osservatorio Astrofisico di Torino - INAF (http://www.oato.inaf.it/blazars/webt/).
\end{acknowledgements}

\bibliographystyle{raa}
\bibliography{bib}

\label{lastpage}

\end{document}